\documentclass[fleqn,12pt,twoside]{article}
\usepackage{espcrc1}

\usepackage{graphicx}
\usepackage{amsmath}
\usepackage{epsfig}

\title{System size dependence of strangeness production at 158 \textit{A}GeV}

\author{\vspace{0.3cm}C.~H\"{o}hne for the NA49 Collaboration}

\begin{document}

\maketitle

{\small \vspace{0.2cm} \noindent
S.V.~Afanasiev$^{9}$,T.~Anticic$^{21}$,
B.~Baatar$^{9}$,D.~Barna$^{5}$, J.~Bartke$^{7}$,
R.A.~Barton$^{3}$, M.~Behler$^{15}$, L.~Betev$^{10}$,
H.~Bia{\l}\-kowska$^{19}$, A.~Billmeier$^{10}$, C.~Blume$^{8}$,
C.O.~Blyth$^{3}$, B.~Boimska$^{19}$, M.~Botje$^{1}$,
J.~Bracinik$^{4}$, R.~Bramm$^{10}$, R.~Brun$^{11}$,
P.~Bun\v{c}i\'{c}$^{10,11}$, V.~Cerny$^{4}$, O.~Chvala$^{17}$,
J.G.~Cramer$^{18}$, P.~Csat\'{o}$^{5}$, P.~Dinkelaker$^{10}$,
V.~Eckardt$^{16}$, P.~Filip$^{16}$, H.G.~Fischer$^{11}$,
Z.~Fodor$^{5}$, P.~Foka$^{8}$, P.~Freund$^{16}$,
V.~Friese$^{8,15}$, J.~G\'{a}l$^{5}$, M.~Ga\'zdzicki$^{10}$,
G.~Georgopoulos$^{2}$, E.~G{\l}adysz$^{7}$, S.~Hegyi$^{5}$,
C.~H\"{o}hne$^{15}$, G.~Igo$^{14}$, P.G.~Jones$^{3}$,
K.~Kadija$^{11,21}$, A.~Karev$^{16}$, V.I.~Kolesnikov$^{9}$,
T.~Kollegger$^{10}$, M.~Kowalski$^{7}$, I.~Kraus$^{8}$,
M.~Kreps$^{4}$, M.~van~Leeuwen$^{1}$, P.~L\'{e}vai$^{5}$,
A.I.~Malakhov$^{9}$, S.~Margetis$^{13}$, C.~Markert$^{8}$,
B.W.~Mayes$^{12}$, G.L.~Melkumov$^{9}$, C.~Meurer$^{10}$,
A.~Mischke$^{8}$, M.~Mitrovski$^{10}$, J.~Moln\'{a}r$^{5}$,
J.M.~Nelson$^{3}$, G.~P\'{a}lla$^{5}$, A.D.~Panagiotou$^{2}$,
K.~Perl$^{20}$, A.~Petridis$^{2}$, M.~Pikna$^{4}$,
L.~Pinsky$^{12}$, F.~P\"{u}hlhofer$^{15}$, J.G.~Reid$^{18}$,
R.~Renfordt$^{10}$, W.~Retyk$^{20}$, C.~Roland$^{6}$,
G.~Roland$^{6}$, A.~Rybicki$^{7}$, T.~Sammer$^{16}$,
A.~Sandoval$^{8}$, H.~Sann$^{8}$, N.~Schmitz$^{16}$,
P.~Seyboth$^{16}$, F.~Sikl\'{e}r$^{5}$, B.~Sitar$^{4}$,
E.~Skrzypczak$^{20}$, G.T.A.~Squier$^{3}$, R.~Stock$^{10}$,
H.~Str\"{o}bele$^{10}$, T.~Susa$^{21}$, I.~Szentp\'{e}tery$^{5}$,
J.~Sziklai$^{5}$, T.A.~Trainor$^{18}$, D.~Varga$^{5}$,
M.~Vassiliou$^{2}$, G.I.~Veres$^{5}$, G.~Vesztergombi$^{5}$,
D.~Vrani\'{c}$^{8}$, S.~Wenig$^{11}$, A.~Wetzler$^{10}$,
C.~Whitten$^{14}$, I.K.~Yoo$^{8,15}$, J.~Zaranek$^{10}$,
J.~Zim\'{a}nyi$^{5}$

\vspace{0.5cm} \noindent
$^{1}$NIKHEF, Amsterdam, Netherlands. \\
$^{2}$Department of Physics, University of Athens, Athens, Greece.\\
$^{3}$Birmingham University, Birmingham, England.\\
$^{4}$Comenius University, Bratislava, Slovakia.\\
$^{5}$KFKI Research Institute for Particle and Nuclear Physics, Budapest, Hungary.\\
$^{6}$MIT, Cambridge, USA.\\
$^{7}$Institute of Nuclear Physics, Cracow, Poland.\\
$^{8}$Gesellschaft f\"{u}r Schwerionenforschung (GSI), Darmstadt, Germany.\\
$^{9}$Joint Institute for Nuclear Research, Dubna, Russia.\\
$^{10}$Fachbereich Physik der Universit\"{a}t, Frankfurt, Germany.\\
$^{11}$CERN, Geneva, Switzerland.\\
$^{12}$University of Houston, Houston, TX, USA.\\
$^{13}$Kent State University, Kent, OH, USA.\\
$^{14}$University of California at Los Angeles, Los Angeles, USA.\\
$^{15}$Fachbereich Physik der Universit\"{a}t, Marburg, Germany.\\
$^{16}$Max-Planck-Institut f\"{u}r Physik, Munich, Germany.\\
$^{17}$Institute of Particle and Nuclear Physics, Charles University, Prague, Czech Republic.\\
$^{18}$Nuclear Physics Laboratory, University of Washington, Seattle, WA, USA.\\
$^{19}$Institute for Nuclear Studies, Warsaw, Poland.\\
$^{20}$Institute for Experimental Physics, University of Warsaw, Warsaw, Poland.\\
$^{21}$Rudjer Boskovic Institute, Zagreb, Croatia.\\  }

\begin{abstract}

{\small Strange particle production in A+A interactions at 158
\textit{A}GeV is studied by the CERN experiment NA49 as a function
of system size and collision geometry. Yields of charged kaons,
$\phi$ and $\Lambda$ are measured and compared to those of pions
in central C+C, Si+Si and centrality-selected Pb+Pb reactions. An
overall increase of relative strangeness production with the size
of the system is observed which does not scale with the number of
participants. Arguing that rescattering of secondaries plays a
minor role in small systems the observed strangeness enhancement
can be related to the space-time density of the primary
nucleon-nucleon collisions.}

\end{abstract}

\section{Motivation}

An increase of the ratio of strange to non-strange particles is
observed in all A+A collisions compared to N+N reactions; it is
found to depend on energy \cite{na49energy}, size of the
interacting nuclei and collision geometry \cite{ferenc}. This
strangeness enhancement was proposed as a signal for a deconfined
state of matter. But despite rich experimental data and many
theoretical efforts the origin of strangeness enhancement is still
not fully understood.

In this contribution we concentrate on the dependence of
strangeness enhancement on the system size and, in particular, on
the collision geometry. Macroscopic, thermodynamic models relate
the increase of strangeness production with system size to the
transition from the canonical to the grand-canonical ensemble
\cite{macro}. The resulting disappearance of what is called
canonical strangeness suppression already in rather small reaction
volumes is based on a mechanism that is still open to discussion.
In a microscopic picture successively excited nucleons create
collision, string and energy densities varying under different
experimental conditions. This provides the opportunity to search
for the microscopic parameters responsible for enhanced
strangeness production.

\section{Experiment and results}

NA49 is a large-acceptance hadron spectrometer \cite{na49nim}.
Charged pions and kaons are identified by means of energy loss and
momentum information. Short-lived particles as the $\phi$ or the
$\Lambda$ are measured through their hadronic decay channels,
i.e.~$\phi\rightarrow K^{+}K^{-}$ and $\Lambda\rightarrow
p\pi^{-}$. For the $\Lambda$ the $V_{0}$-decay topology is used in
addition. All data are corrected for acceptance, kaon decay in
flight and the vertex resolution.

The data on p+p and centrality-selected Pb+Pb interactions were
presented elsewhere
\cite{na49energy,ferenc,na49phi,volker,tanja,andre}. The light
systems C+C and Si+Si were investigated using a secondary beam of
Pb-fragments (for further details see \cite{na49nim}). Central
collisions were chosen by setting an upper threshold on the
forward energy as measured in the zero-degree calorimeter. The
percentage of the inelastic cross section selected this way was
used to calculate the mean number of participants $N_{part}$ as
well as the mean number of collisions $\nu$ within the VENUS model
(version 4.12) \cite{venus}. In C+C (Si+Si) the 17.5\% $\pm$ 1.5\%
(12.5\% $\pm$ 1.5\%) most central events correspond to 16 $\pm$ 1
(41.5 $\pm$ 1.5) participants which undergo 1.7 (2.2) collisions
on average.

\begin{figure}[htb]
\vspace*{-0.8cm}
\begin{minipage}[t]{5.7cm}
 \begin{center}
  \epsfig{file=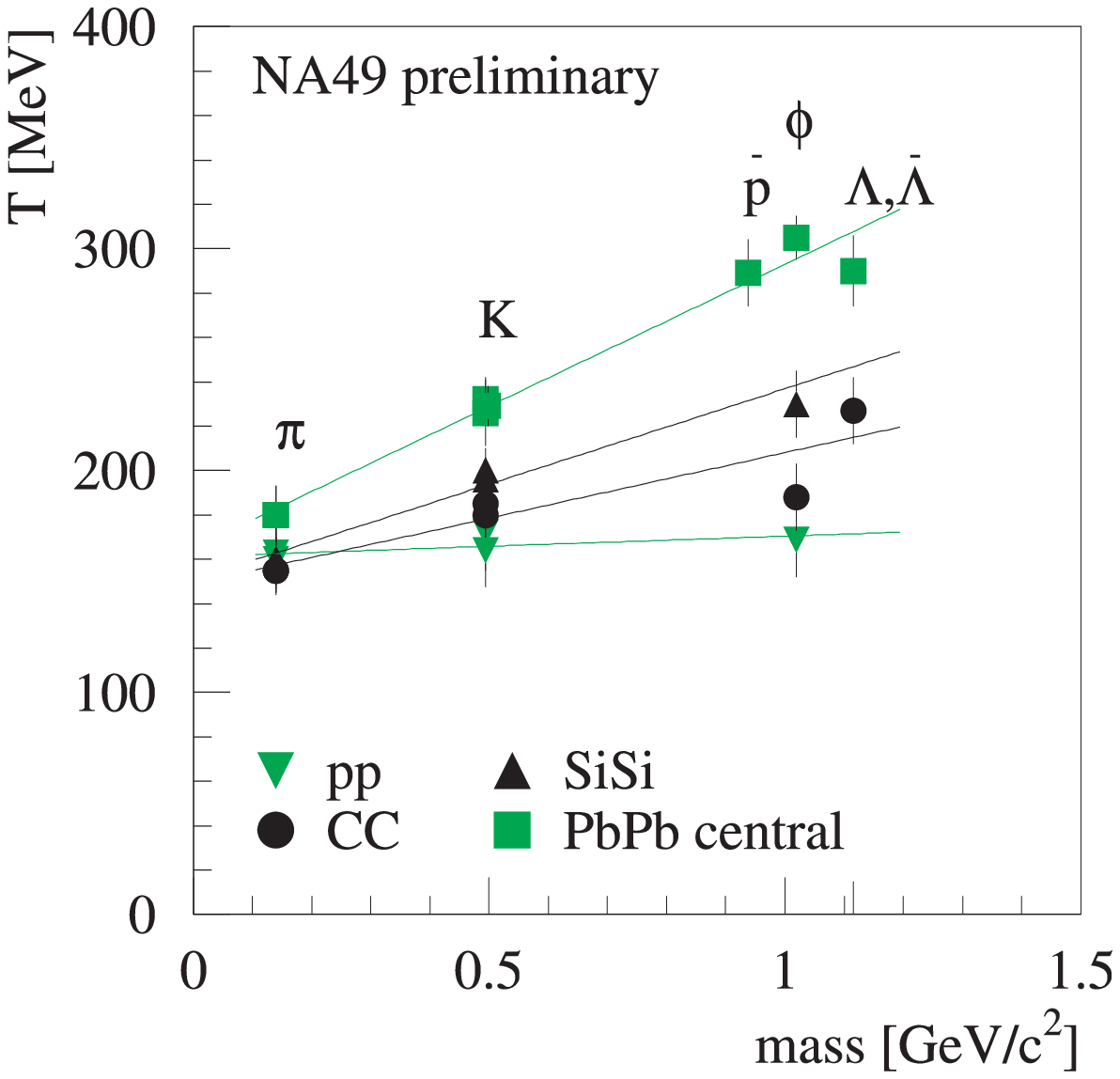,width=5.7cm}
 \end{center}
 \vspace*{-1.2cm}
 {\caption{{\small Inverse transverse slope parameters as a function of particle
 mass and system size.} \label{mt}}}
  \end{minipage}
  \hspace{\fill}
\begin{minipage}[t]{9.6cm}
 \begin{center}
  \epsfig{file=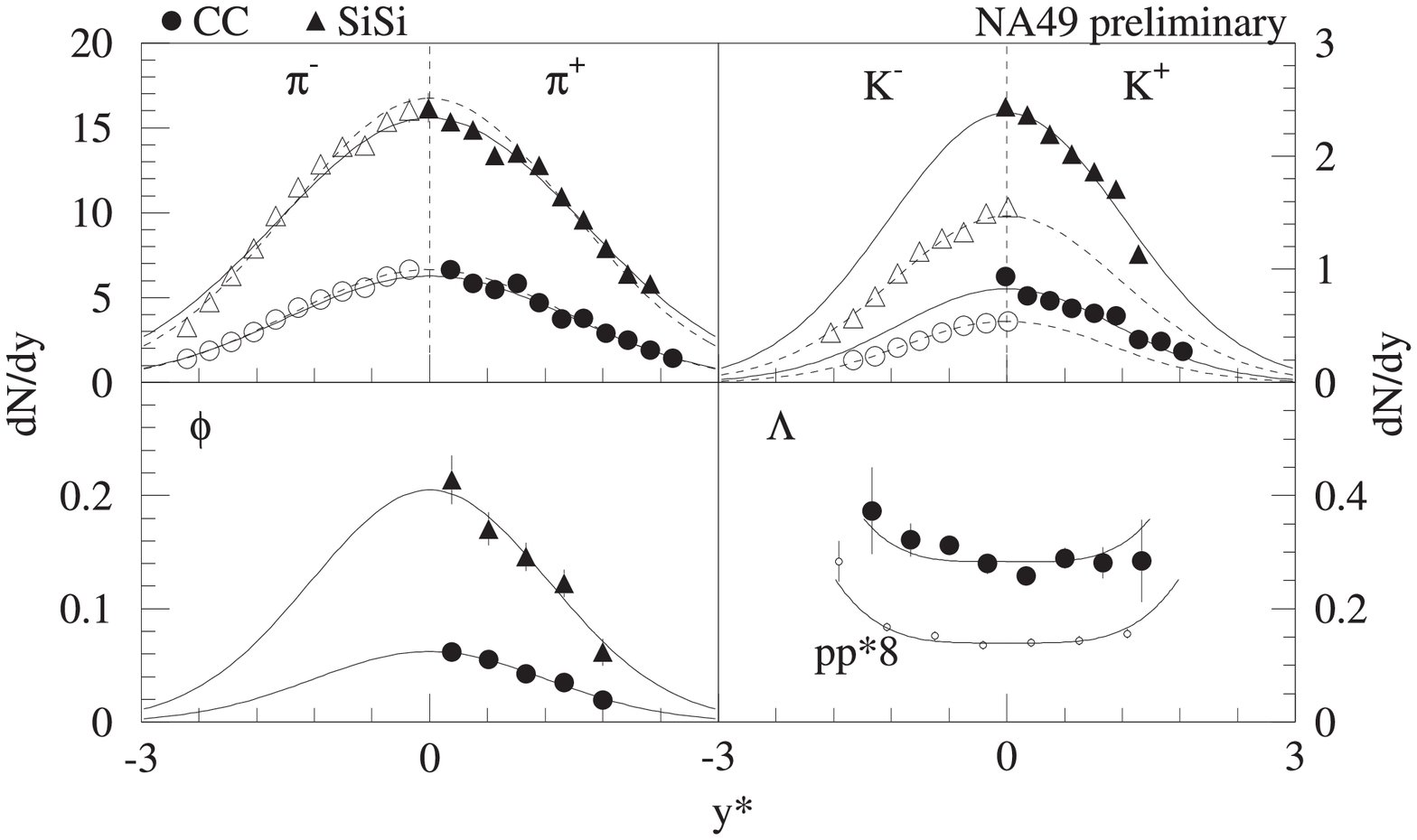,width=9.6cm}
 \end{center}
  \vspace*{-1.2cm}
 {\caption{{\small Rapidity distributions for C+C and Si+Si; for clearness, negative
 particles are plotted as open symbols at reflected y-positions. The $y$-distribution of $\Lambda$ in p+p
 \cite{tanja} scaled by $8=N_{part}(\text{CC})/N_{part}(\text{pp})$ is also
 shown, lines are to guide the eye.} \label{y}}}
  \end{minipage}
\vspace*{-0.8cm}
\end{figure}

Transverse mass distributions in central C+C and Si+Si
interactions show an exponential shape. The mass dependence of the
inverse slopes (fig.~\ref{mt}) indicates the presence of
collective transverse flow which itself increases with system
size. Rapidity spectra of mesons (fig.~\ref{y}) are fitted by a
Gaussian to extract full yields. In the figures only statistical
errors are shown; additional systematic errors for the yields
amount to about 5\% (10\%) for pions (kaons, $\phi$). The rapidity
distribution of the $\Lambda$ is rather flat over the whole
measured range. A comparison to scaled p+p data shows an increase
on the order of two at midrapidity which reflects both, larger
stopping and strangeness enhancement.

\section{Discussion and Interpretation}

\begin{figure}[t]
\vspace*{-0.cm}
\begin{minipage}[t]{11.7cm}
 \begin{center}
  \epsfig{file=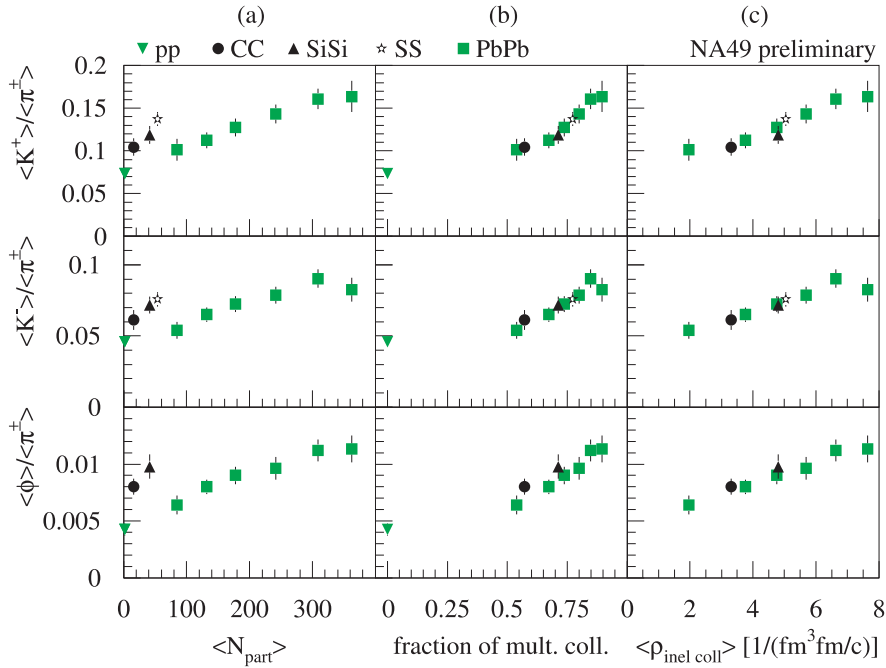,width=11.7cm}
 \end{center}
 \end{minipage}
  \hspace{\fill}
\begin{minipage}[htb]{3.8cm}
 \vspace*{-8.6cm}
 {\caption{{\small Strange hadron production relative to pions
 ($\langle\pi^{\pm}\rangle=(\langle\pi^{+}\rangle+\langle\pi^{-}\rangle)/2$)
 in dependence on parameters as defined in the text. Ratios for S+S collisions are from
 \cite{na35}, those for centrality selected Pb+Pb reactions from \cite{na49energy,ferenc,na49phi,volker}.} \label{npart}}}
 \end{minipage}
\vspace*{-1cm}
\end{figure}

Relative strangeness production in nuclear collisions is
approximately measured by the $\langle K\rangle/\langle\pi\rangle$
ratio, since roughly 70\% of the produced strangeness appears in
the kaons. The $\phi$-meson is of interest because of its hidden
strangeness. Ratios of these particles are shown in figure
\ref{npart}(a) as function of the number of participants.
Strangeness enhancement relative to p+p interactions which
increases with system size is already observed in small systems.
However, the ratios are higher in central collisions of small
systems than in peripheral Pb+Pb at the same number of
participants indicating an important effect of the collision
geometry on strangeness production. Previously \cite{christoph}
this was parametrized with the macroscopic geometrical variable
$R-b/2$ representing the surface per volume ratio of the system.
Here, this geometry effect will be discussed in terms of a
microscopic reaction picture: The strategy is to search for a
common scaling parameter which should give insight into the
underlying reaction mechanism.

In e$^{+}$+e$^{-}$ and p+p collisions an approximate
energy-independence of the $\langle K\rangle/\langle\pi\rangle$
ratio is observed in the range of interest here
\cite{na49energy,pdg}. This suggests that, aside from at most a
few percent, higher excitation of nucleons by sequential N+N
interactions in A+A collisions alone does not cause the
strangeness enhancement.

Another new feature in A+A reactions compared to p+p is
rescattering of secondaries. An indication that this can be
neglected in light systems comes from the small collective
transverse flow observed in C+C and Si+Si in comparison to central
Pb+Pb collisions (fig.~\ref{mt}). In addition UrQMD\footnote{The
author thanks the UrQMD collaboration for providing the
preliminary version 1.3.} \cite{urqmd} shows that the increase of
the $\langle K\rangle/\langle\pi\rangle$ ratio from rescattering
is minor in small systems, and, more importantly, that more
rescattering takes place in peripheral Pb+Pb than in central C+C
and Si+Si reactions. This is plausible because of the smaller
particle multiplicity in the latter systems. Thus, if rescattering
would be the source of strangeness enhancement peripheral Pb+Pb
should give larger ratios.

Assuming that energy loss and rescattering are not the main
sources of strangeness enhancement other features of the primary
inelastic N+N collisions must play an important role. A first
obvious attempt is to investigate the role of multiple
interactions. As shown in \cite{ferenc}, the mean number of
collisions per projectile $\nu$ as calculated within the Glauber
approach does not provide scaling; this holds also for C+C and
Si+Si collisions. However, the distribution of $\nu$ for C+C or
Si+Si and Pb+Pb reactions at the same $\langle
K\rangle/\langle\pi\rangle$ ratio shows that the fraction of
nucleons which undergo multiple collisions is similar. In fact
scaling is observed in this latter variable (fig.~\ref{npart}(b)).

The sequential N+N reactions happen in close vicinity to each
other, thus causing a high density of interactions in space and
time. Indeed, the mean space-time density of all inelastic
collisions during the first phase $\langle\rho_{\text{inel\
coll}}\rangle$, i.e.~when the nucleons pass through each other, as
calculated in the center of mass system of the collision within
the UrQMD model (version 1.3) serves also as scaling parameter
(fig.~\ref{npart}(c)).

\section{Conclusions}

The presented data show that the number of participating nucleons
in A+A collisions is not the decisive variable for strangeness
enhancement (fig.~\ref{npart}(a)). However, taking advantage of
this observation it is found that features connected to the
primary inelastic N+N collisions provide a common description of
the $\langle K^{+}\rangle/\langle\pi^{\pm}\rangle$, $\langle
K^{-}\rangle/\langle\pi^{\pm}\rangle$ and the $\langle
\phi\rangle/\langle\pi^{\pm}\rangle$ ratio in various A+A
collision systems. This implies that sequential N+N interactions
within a small volume and period of time are not independent of
each other with regard to strangeness production.

The collision density as defined here is on the one hand related
to the energy density in the system. On the other hand a high
density of interactions may also mean overlap of strings thus
leading to string fusion and enhanced string tension. Both, string
overlap and energy density have often been discussed as being
relevant for enhanced strangeness production. In addition a volume
in which a high energy density or overlapping strings and thus a
high space-time density of successive collisions exist might decay
in a quantum-mechanically coherent fashion, and thus represent the
hadronizing volume implied by the canonical and grand-canonical
versions of the statistical hadronization model.

\vspace*{0.15cm}

{\footnotesize Acknowledgements: This work was supported by the
Director, Office of Energy Research, Division of Nuclear Physics
of the Office of High Energy and Nuclear Physics of the US
Department of Energy (DE-ACO3-76SFOOO98 and DE-FG02-91ER40609),
the US National Science Foundation, the Bundesministerium f\"{u}r
Bildung und Forschung, Germany, the Alexander von Humboldt
Foundation, the UK Engineering and Physical Sciences Research
Council, the Polish State Committee for Scientific Research (2
P03B 130 23 and 2 P03B 02418), the Hungarian Scientific Research
Foundation (T14920 and T32293), Hungarian National Science
Foundation, OTKA, (F034707), the EC Marie Curie Foundation, the
Polish-German Foundation, and Bergen Computational Physics
Laboratory in the framework of the European Community - Access to
Research Infrastructure action of the Improving Human Potential
Programme.}

\end{document}